\documentclass[11pt, oneside]{article}
\usepackage{authblk}
\usepackage{geometry}
\usepackage{graphicx}
\usepackage{amssymb}
\usepackage{amsmath}
\usepackage{url}
\usepackage{comment}
\usepackage{booktabs}
\usepackage{setspace}
\usepackage[square, numbers]{natbib}
\usepackage{lineno}
\PassOptionsToPackage{hyphens}{url}
\usepackage{hyperref}

\makeatletter

\author[2]{Maria Cuellar}

\affil[2]{University of Pennsylvania Department of Criminology and Statistics}

\title{The Neglected Error: False Negatives and the Case for Validating Eliminations}

\begin{document}

\maketitle

\begin{abstract}
This article examines the overlooked risk of false negative errors arising from eliminations in forensic firearm comparisons. While recent reforms in forensic science have focused on reducing false positives, eliminations—often based on class characteristics or intuitive judgments—receive little empirical scrutiny despite their potential to exclude true sources. In cases involving a closed pool of suspects, eliminations can function as de facto identifications, introducing serious risk of error. A review of existing validity studies reveals that many report only false positive rates, failing to provide a complete assessment of method accuracy. This asymmetry is reinforced by professional guidelines, such as those from AFTE, and echoed in major government reports, including those from NAS and PCAST. The article argues that eliminations, like identifications, must be validated through rigorous testing and reported with transparent error rates. It further cautions against the use of “common sense” eliminations in the absence of empirical support and highlights the dangers of contextual bias when examiners are aware of investigative constraints. Five policy recommendations are proposed to improve the scientific treatment and legal interpretation of eliminations, including balanced reporting of false positive and false negative rates, validation of intuitive judgments, and clear warnings against using eliminations to infer guilt in closed-pool scenarios. Without reform, eliminations will continue to escape scrutiny, perpetuating unmeasured error and undermining the integrity of forensic conclusions.
\end{abstract}


\section{Introduction}

The field of forensic firearm comparisons has undergone significant reform efforts in recent years, driven by concerns about reliability and error rates. However, these efforts have overwhelmingly focused on reducing false-positive errors, with little attention paid to the risks of incorrect eliminations. Eliminations, when made without sufficient empirical basis, can lead to false-negative errors that result in miscarriages of justice. They can also lead to false-positive errors when paired with a closed pool of suspects.

This article examines the overlooked problem—and downstream consequences—of prioritizing false-positive errors over false-negative ones in the criminal justice system, particularly in forensic disciplines where examiners are permitted to make eliminations without adequate validation. It begins by outlining the field’s narrow focus on false positives, then shows that many validity studies in firearm comparisons fail to report false negative rates. It argues that this imbalance may stem from normative assumptions in the law and demonstrates that even major government reform efforts, such as the NAS and PCAST reports, reflect this same bias. The article further highlights that eliminations based on “common sense” have not been empirically studied, that AFTE’s guidelines for eliminations can disadvantage defendants, and that eliminations can function as de facto identifications when a closed pool of contributors is presumed. A formal hypothetical example illustrates the risks of this logic. The article concludes with five policy recommendations aimed at improving the scientific reliability and legal interpretation of eliminations in forensic firearm analysis.

\section{There is a myopic focus on false positives}

In the field of firearms and toolmarks, examiners give their conclusions about whether two fired bullets or cartridge casings came from the same source as one of three categories: identification, elimination, or inconclusive. This originated in 1992, when the Association of Firearm and Toolmark Examiners (AFTE) adopted the Range of Conclusions \cite{AFTE1992Criteria}, which is now called the ``AFTE Theory of Identification'' \cite{aftetheory}.

Errors in conclusions can come as a false-positive (FP) error, when an examiner incorrectly classifies two different-source samples as an identification, and a false-negative (FN) error, when an examiner incorrectly classifies two same-source samples as an elimination. When the method correctly identifies a same-source pair, this is a true positive (TP), and when it correctly identifies a different-source pair, this is a true negative (TN). A method should have a properly measured false positive rate (FPR) and false negative rate (FNR),
\begin{align}
\text{FPR} & = \frac{\text{FP}}{\text{FP+TN}}\\
\text{FNR} & = \frac{\text{FN}}{\text{FN+TP}}.
\end{align}
One can summarize a method's performance is by using the measures of sensitivity and specificity,
\begin{align}
\text{Sensitivity} & = \frac{\text{TP}}{\text{TP+FN}}\\
\text{Specificity} & = \frac{\text{TN}}{\text{TN+FP}}.
\end{align}

Sensitivity can be understood as the method's ability to identify correctly same-source bullets or cartridges. If one takes a car alarm as a method to detect when someone is breaking into a car, an alarm that is too sensitive will go off when a bird steps on the car, or when the wind blows. This alarm is not useful because it goes off so often that one does not trust it anymore. Specificity can be understood as the method's ability to correctly eliminate different-source bullets or cartridges. Following the car analogy, an alarm that is too specific is quiet too often. It will sometimes not go off even if someone breaks into the car because it is tuned so narrowly that only very precise, rare events will trigger it. This kind of alarm might avoid false alarms, but it fails in its primary purpose: detecting real break-ins. Similarly, in forensic firearm examination, a method that is overly specific may miss true matches because it is too cautious about declaring a match. An ideal method balances sensitivity and specificity to maximize overall accuracy in distinguishing between same-source and different-source comparisons.

The relationship between FPR/FNR and sensitivity/specificity is, 
\begin{align}
\text{Sensitivity} & = 1-\text{FNR}\\
\text{Specificity} & = 1-\text{FPR}.
\end{align}

Thus, to evaluate a method, it is essential that the researchers include both the FPR \textit{and} the FNR (or, the sensitivity \textit{and} specificity), because one without the other gives an incomplete picture \cite{pcast}. To understand why this is so, it is useful to think about an extreme case. If only the false positive error rate were of concern, then an examiner could conclude that every pair of bullets was an elimination. This would yield a zero percent false positive rate, but it would do so at the expense of having a useless method. In other words, it's always possible to set a false positive rate to zero if one ignores the false negative rate, or vice versa.

\section{Validity studies of firearms comparisons often do not report a false negative rate}

Since the National Academies Report entitled ``Strengthening forensic science in the United States: A path forward'' was published in 2009, several black-box studies have been performed to evaluate the validity of forensic firearm comparisons. \cite{cuellar2024methodological} found 28 such studies. It is a worthwhile exercise to check whether the existing validity studies have this bias towards only taking false positive errors into account.

After reviewing the 28 studies, I found that only 45\% of validity studies of firearms comparisons (as selected by \cite{cuellar2024methodological}) report both FPR and FNR, or sensitivity and specificity. 20\% of the 28 validity studies available for firearm comparisons do not cite a method’s false negative rate because they do not split up the errors into FPR and FNR. And 35\% do not report errors or have no errors. This suggests that the study design for this last group was not adequate.

\section{This myopic focus might arise from normative foundations of the law}

Perhaps this bias arises from the normative underpinning of our legal system. Justice Blackstone once famously wrote: ``It is better that ten guilty persons escape than that one innocent suffer.'' \cite{lieberman1988blackstone} The question of whether the specific ratio in Blackstone’s formulation is the correct one to use has inspired debate for more than two centuries \cite{cole2005does}. Its underlying principle serves as a core value throughout the Western legal tradition. Blind adherence to it can obscure errors in eliminations. In other words, simply worrying about false positives, which could lead to a determination of guilt for an innocent person, threatens to ignore errors of elimination.

In fingerprint comparisons, the case of United States v. Mitchell, 365 F.3d 215 (3d Cir. 2004), illustrates this bias \cite{cole2005does}. In that case, judges were explicit that false positive errors is the type of error that matters. This reasoning suggests that in pattern-matching disciplines, false positives are the most consequential type of error.

One might think that examiners seek to minimize false positive errors, and although it's true in validity studies \cite{hofmann2020treatment}, we do not have evidence to say that this it true in casework. There is evidence that examiners do not set thresholds to minimize false positive errors \cite{Scurich2024}. Even assuming that examiners do seek to avoid false positive errors in studies, we cannot accept that as proof that they do so in casework because the frequency of inconclusives is far greater in these studies than casework \cite{scurich2025hawthorne}. These findings strengthen the concerns that the AFTE Range of Conclusions leads firearm examiners to make inconsistent decisions. 

Although it might be normatively worse to incriminate an innocent individual in a court of law than to let a guilty person go free, it is nevertheless important to consider both types of errors when evaluating the scientific validity of a forensic science method.

\section{PCAST and NAS suffer from this myopic focus too}

The President’s Council of Advisors on Science and Technology (PCAST) 2016 report, ``Forensic Science in Criminal Courts: Ensuring Scientific Validity of Feature-Comparison Methods'' \cite{pcast}, is aware of the bias towards worrying about false positives. The report states that, ``Under some circumstances, false-negative results can contribute to wrongful convictions as well,'' \citep{pcast} (p.50) and thus there are grave consequences to having a false negative error. 

\begin{figure}[h]
   \centering
   \includegraphics[width=\textwidth]{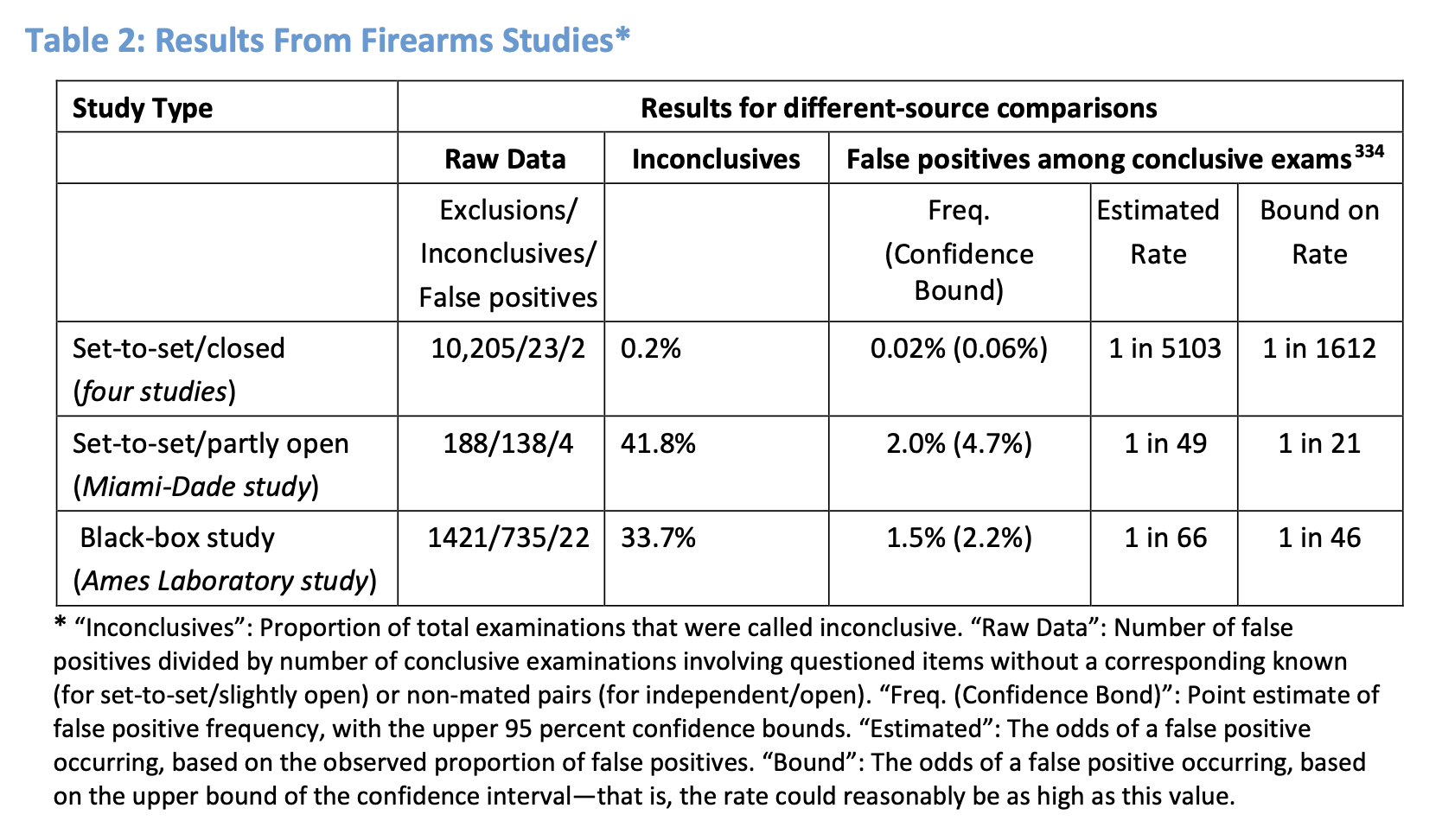} 
   \caption{Table from the PCAST report showing only the number of false positive errors in studies.}
   \label{fig:pcasttable2}
\end{figure}

Nevertheless, PCAST suffers from this bias as well. \citep[p.~110]{pcast} only reports the number of false positive errors in the study that they think is properly designed. When critiquing other study designs (closed and partly open), PCAST provides a table shown here as Figure \ref{fig:pcasttable2}, but it only provides the number of false positive errors.

PCAST states that in a 2009 article \cite{Bunch2009}, the chief of the firearms-toolmarks unit of the FBI Laboratory stated that ``a qualified examiner will rarely if ever commit a false-positive error (misidentification),'' citing his review, in an affidavit, of empirical studies that showed virtually no errors. Although PCAST provided this quote to argue that the firearms discipline is not sufficiently concerned about errors, note that the chief of the FBI Laboratory focuses \textit{only on whether examiners make false-positive errors}.

Both the NAS 2009 report and the PCAST 2016 report both allow for eliminations to be made with less supportive evidence than identifications. This fault appears in other pattern-matching disciplines as well. For example, \citep[p.~89]{pcast} states that ``Fingerprint features are compared at three levels of detail-level [1,2,3]. `Ridge flow' refers to classes of pattern types shared by many individuals, such as loop or whorl formations; \textit{this level is only sufficient for eliminations, not for declaring identifications.}'' And \citep[p.~176]{nas2009} states that, ``Despite the inherent weaknesses involved in bite mark comparison, \textit{it is reasonable to assume that the process can sometimes reliably exclude suspects}.'' (Italics added.) For PCAST's and NAS's recommendations to actually strengthen forensic science, research on the validity of the methods needs to focus on both false positive and false negative errors.

\section{Eliminations based on common-sense information have not been sufficiently studied}

Evaluating the difficulty of a firearms comparison requires accounting for the nature of the comparison—specifically, whether it involves a close non-match or a far match. Close non-matches occur when two bullets or cartridge cases from different sources appear highly similar, making them difficult to distinguish and increasing the risk of false positives. Far matches, on the other hand, involve same-source items that appear dissimilar due to damage, poor-quality markings, or other distortions, increasing the risk of false negatives. 

\begin{figure}[h]
   \centering
   \includegraphics[width=.7\textwidth]{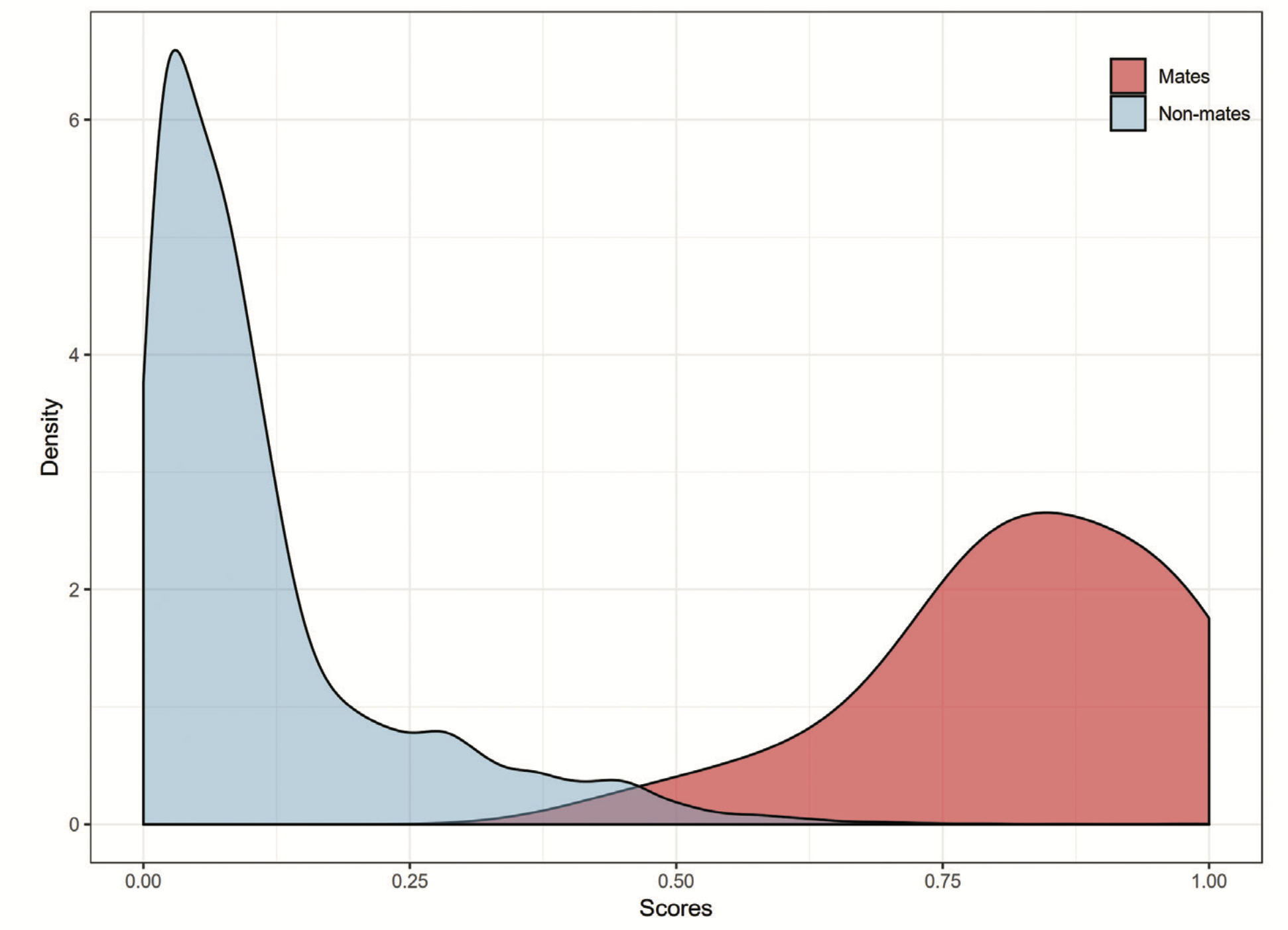} 
   \caption{Densities of the similarity scores from the known same-source pairs (``mates'') and the known different-source pairs (``non-mates'')  from \cite{carriquiry2019machine}. Even after defining a similarity score, the densities are not perfectly separated and have some overlap.}
   \label{fig:densities}
\end{figure}

Figure \ref{fig:densities} from \cite{carriquiry2019machine} shows an example of close-non-matches and far-matches in training densities for an algorithm. Even after defining a similarity score using machine learning, which can classify with very high performance, the densities are not perfectly separated and have some overlap. It is likely that ``difficult'' cases like these (in the overlap) appear in court, and thus errors in comparisons are more likely. 

These cases represent the most challenging comparisons and contribute disproportionately to examiner errors. An objective measure of difficulty should incorporate how visually confusable the items are—whether due to misleading similarity or unexpected dissimilarity—so that we can provide more accurate, context-specific error rates. This framework would also allow courts to distinguish between cases that genuinely require expert analysis and those where the comparison is so clear-cut that common sense may suffice.

We can imagine that any rational person could determine whether two samples are in the same class or not with near perfect accuracy. So, for well-defined, clear, and consistent (class or other) characteristics, eliminations based on these are likely a small source of error. Nevertheless, errors of classification across class characteristics do happen.

Many view eliminations based on class characteristics as ``common sense'', and assume that trained examiners can make these reliably without committing errors. Class characteristics are defined as design-based features established prior to manufacture that indicate a limited group source. In firearms, these include the size, shape, and orientation of parts. Individual characteristics, by contrast, are unique marks resulting from random imperfections caused during manufacturing, use, or damage, and they serve to distinguish one tool from all others. Subclass characteristics fall between the two: they are shared by tools produced by the same manufacturing process and may resemble individual characteristics, but they only reflect a smaller subset within a class. \cite{nij_physical_characteristics}.

This assumption, however, is not true. There is emerging evidence that trained firearms examiners cannot make out-of-class eliminations reliably, although more research is needed to determine the frequency of these errors. For example, \cite{monson2023} found that 1/3 of all eliminations were reportedly based on class characteristics despite the study having no out-of-class comparisons, and \cite{hicklin2024} found that trained examiners had issues with out-of-class eliminations.

Although class characteristic misidentifications in firearms analysis are rarely documented in real-world casework, one known example emerged from Rhode Island and was discovered through anomalies in Collaborative Testing Services: Forensics Testing Program (CTS) proficiency test 23-5262 \cite{cts}. That case highlights the potential for misclassification at the level of class characteristics. Related concerns have also been raised in the research literature. For instance, Scurich and colleagues discussed disagreements among examiners in the FBI's black box study \cite{amesii} about whether eliminations could be based solely on class characteristics \cite{ScurichStern2023}. In their commentary, they highlighted that some examiners made elimination decisions based on class characteristics, despite the study's design intending to preclude such conclusions. This discrepancy underscores variability in examiner interpretations and the need for clearer guidelines in forensic firearm examinations

Similarly, \cite{BestGardner2022} reported that in one test item designed to be a class characteristic elimination, 29 out of 67 participants (43\%) failed to exclude, suggesting uncertainty or disagreement about class-level features. Additional evidence of this challenge comes from comparisons involving polygonal rifling, where consistent classification has proven difficult (see \cite{Hocherman2003}). Ongoing work is examining when and how disagreements about class characteristics appear in CTS proficiency test results. One striking example comes from CTS 23-5261, in which 80 examiners who described their eliminations cited class-level differences, while 56 stated that the class characteristics were the same but individual characteristics differed—revealing substantial disagreement over the basic classification of the marks. It is important to have an empirically determined measure of error for eliminations based on class characteristics.

We might be tempted to think that in these situations, the examiner should be allowed to make eliminations based on class characteristics since these are ``easy cases''. Because we do not have a standardized well-defined measure of how easy or difficult a case is, a case that might seem easy to one person might be complex to another. This is a \textbf{slippery slope} because even if there are in fact ``easy cases'', and we allow common sense to be used in deciding those, more and more complex cases could be considered easy, and this could lead to errors.

However, in the cases in which eliminations seem self-evident, the expert should focus on presenting the ``raw'' pattern information rather than offering their interpretation. This allows the trier of fact to rely on their own understanding of the evidence to form their own sense of uncertainty. However, when the expert believes their interpretation adds value beyond simply presenting the raw data—such as when expertise is necessary to interpret the evidence—there should be data available to quantify the value of the expert’s opinion. 

To summarize, it may be intuitively true that people can make eliminations based on common sense information. However, insofar as this is used in a court, or cited as an expert opinion, it must be subjected to the same rigorous testing that less intuitive approaches receive. Just because an elimination seems obvious, this is not a reason for that type of claim to go untested.

\section{The AFTE guidelines are biased against the defendant}

The AFTE Theory of Identification \citep{aftetheory} guidelines on eliminations provide an interesting case of a bias against false negative errors. In other words, AFTE makes it  more challenging for an examiner to select an elimination than an identification, based on ``individual'' characteristics, when class-chracteristics agree. These are the guidelines given by AFTE,
\begin{quote}
For purposes of fired cartridge case and shotshell case comparisons, an elimination is most often based on observed differences in any class characteristic.

However, an elimination based on individual characteristics is more complex. Conceptually, an elimination based on individual characteristics means that if a firearm can be shown to have never been subjected to significant use or abuse over a period of time, the qualitative aspects of the striations produced on fired cartridge cases and shotshell cases should remain the same. A difference in these qualities indicates an elimination.

Elimination based on individual characteristics requires a detailed knowledge of the history and treatment of the firearm, as well as documentation to support the history. It is the responsibility of the examiner to provide this historical documentation. This type of elimination should be approached with caution. Many experienced examiners have never made such an elimination, and the protocols of many laboratories do not allow it.
\end{quote}

AFTE warns against making an elimination based on ``individual'' characteristics, without having more information. In other words, this policy states that eliminations should not be made when class characteristics agree. Some laboratories have a policy prohibiting out-of-class eliminations by policy. For example, the FBI’s firearm and tool mark laboratory prohibits such class-based eliminations and requires that these comparisons be deemed inconclusive. %CITE?

Firearm examiners often will not make eliminations (when class characteristics agrees) without information about when the crime scene samples were fired, examination of the firearm, etc. They assert that this information could help explain the absence of agreement in individual characteristics, demonstrating their concern with false negatives in casework. The concern with this is whether there is contextual bias from the additional information provided. If additional information includes task-irrelevant information, there is a risk in making errors in same-class cases.

AFTE's guidelines are often biased against defendants, in cases in which the an identification incriminates the defendant. The warning that eliminations should be made with more information should state that the information needs to be task-relevant, and should also apply to identifications.

\section{From elimination to incrimination: When false negatives imply false positives}

In cases where investigators have identified a closed pool of potential firearms—such as when a limited number of suspects and their weapons are known, eliminations can play a central role in implicating a particular firearm. If forensic examiners determine that all but one of the firearms can be excluded as the source of a bullet or cartridge case, the remaining firearm may be inferred to be the source by process of elimination. This inference can occur even if the evidence lacks sufficient individual characteristics to warrant a definitive identification. As described in \cite{bunch2009match}, this logic can effectively transform a series of eliminations into an implicit inclusion, despite the absence of direct individualizing evidence. This is logically valid. As Sherlock Holmes states, ``When you have eliminated the impossible, whatever remains, however improbable, must be the truth.'' But, it is sound only if the members of the pool have been correctly specified. Such use of eliminations in a closed set underscores the importance of understanding how forensic conclusions are interpreted in investigative and legal contexts.

A common but under-examined inferential error arises when forensic examiners or decision-makers treat the elimination of all but one candidate from a closed set as evidence that the remaining individual must be the true source. This form of elimination-based reasoning presumes that the true source is present in the candidate set, and it treats the absence of disqualifying evidence as equivalent to positive identification. Such reasoning is particularly concerning in database-driven forensic searches (e.g., AFIS or NIBIN), where examiners may be presented with a ranked list of candidates and infer that the top-ranking individual is the correct match by process of elimination, even if the match quality is poor or the differences between candidates are minimal.

This inferential pattern is closely related to well-documented phenomena in eyewitness identification research, particularly relative judgment bias, in which witnesses choose the person who appears most similar to the perpetrator in a simultaneous lineup, regardless of whether any lineup member is an actual match \cite{Wells1993}. It also aligns with the guilty suspect assumption in lineup construction, wherein it is erroneously presumed that the suspect is among those presented. Despite its potential to undermine the integrity of forensic conclusions, this mode of reasoning has not been explicitly named or systematically studied within forensic science, and its implications for error rates and validation practices warrant further investigation.

When the police investigation has determined there is a closed pool of candidate sources (or suspects), eliminations are used as a way to incriminate an individual, by process of elimination. If the pool of suspects could have been selected incorrectly, the uncertainty with that decision should be quantified. 

The selection of a ``pool'' is outside the purview of the examiner, but the elimination is used to reach an identification by other actors in the process. How complete is the set of hypotheses to be considered? How accurate is the selection of the ``pool'' of suspects? Without answers to these questions, the conclusion to the compound hypothesis (firearms comparison plus closed pool of sources) could lead to false positive errors. The forensic examiner should not know that the investigators have a closed pool or the size of the pool. The forensic examiner should simply be asked to analyze the evidence without any context. The forensic examiner should remain context-free \cite{dror2015context}. They should warn investigators that they should not use a process of elimination to determine an identification. An inclusion in the pool of potential contributors is not an identification.

This is an example of the black-and-white fallacy, also known as a false dilemma or false dichotomy. That is a situation in which an individual presents fewer choices than there actually are, and through process of elimination selects the wrong choice. The reality is that there is some uncertainty and error in the selection of the pool of potential contributors, and that needs to be taken into account in the way the evidence is presented to the trier of fact, as well as in the trier of fact's final determination of guilt.

\section{Inconclusives}

Finally, inconclusives are relevant to error rates. Combining the focus on false positives and the way in which inconclusives are included in error rate calculations could result in error rates that are artificially low. If an examiner classifies a sample as inconclusive, there is no consensus about whether this should be considered an error or not, and this lack of standardization has led to uneven calculations of error rates in black box studies \cite{hofmann2020treatment}. There is an ongoing debate about how inconclusives should be accounted for when measuring the accuracy of a method \cite{hofmann2020treatment, swofford2024inconclusive}. While a discussion of this debate is outside the scope of this article, one point does bear on eliminations. As Scurich says in his 2022 commentary \cite{scurich2022inconclusives}, inconclusives in firearm error rate studies are not `a pass'. Scurich argues that the common practice of treating inconclusive responses in firearm error rate studies as neutral or non-errors biases reported error rates and undermines their credibility. Classifying a non-identification as an inconclusive rather than an elimination could arbitrarily lower the real false negative rate.

\section{Policy recommendations to improve the validity of eliminations}

In light of the foregoing discussion, I provide five recommendations for practice and policy.

\begin{enumerate}
    \item \textbf{Validity studies}: Just because there are more validity studies than there used to be, not all report their false positive rate \textit{and} their false negative rate (indeed, only 45\% of the existing firearms validity studies report both FPR and FNR). There is a focus on false positive rates, and only mentioning false positive rates without knowing the false negative rate gives an incomplete and useless picture.
    \item \textbf{Common-sense conclusions}: Eliminations based on “common sense” should not be exempt from empirical validation, as intuitive judgments can still lead to error. Objective measures of difficulty must be developed and tested to ensure that even seemingly obvious conclusions meet scientific standards.
    \item \textbf{AFTE guidelines for eliminations}: AFTE's guideline that a within-class elimination should not be made based on ``individual'' characteristics is biased against defendants, in cases in which the an identification incriminates an individual. The warning should apply to identifications. Furthermore, AFTE's recommendation that eliminations should be made with more information should state that the information needs to be task-relevant.
    \item \textbf{Closed pool warning}: The forensic examiner should warn investigators that they should not use a process of elimination to determine an identification. An inclusion in the pool of potential contributors is not an identification.
    \item \textbf{Inconclusives}: The way that inconclusives are included into error rates in validity studies matters. The complex debate about inconclusives is outside the scope of this article. Nevertheless, it is relevant to note that if there is a focus only on false positives, inconclusives could be treated as false negative errors, a practice that would not give an honest or unbiased assessment of the error rates.

\end{enumerate}

\section{Conclusion}

Eliminations in forensic firearm comparisons are frequently assumed to be straightforward, objective, and error-free—especially when based on class characteristics or when the evidence appears intuitively obvious. Yet, as this article has shown, such assumptions are unfounded without rigorous empirical validation. The forensic field has disproportionately focused on avoiding false positive errors, neglecting the very real risk of false negatives that can arise from unsupported eliminations. This imbalance has distorted research priorities, examiner training, and legal interpretations of forensic evidence. In contexts where investigators presume a closed pool of suspects, eliminations are especially dangerous: they can be misinterpreted as positive identifications, thereby producing unjust outcomes without meeting the evidentiary standards required for inclusion.

To ensure that eliminations are treated with the same scientific rigor as identifications, forensic methods must be validated for both types of conclusions, with clearly reported false positive and false negative error rates. Assumptions about ease or ``common sense'' accuracy should never substitute for tested reliability. Furthermore, reporting practices in research, as well as protocols for testimony and investigation, must be updated to account for the role eliminations play in shaping downstream inferences in legal proceedings. Without such reforms, the criminal justice system will remain vulnerable to misinterpretations of forensic conclusions—particularly those that carry the appearance of certainty but are built on a foundation of unmeasured error. Although it might be normatively worse to incriminate an innocent individual in a court of law than to let a guilty person go free, as Justice Blackstone said, it is important to consider both types of errors when evaluating the scientific validity of a forensic science method. Ultimately, eliminations should receive the same rigorous scrutiny as identifications to avoid errors.

% Bibliography
\bibliographystyle{plainnat}
\bibliography{bibeliminations.bib}

\end{document}